\documentclass[runningheads]{llncs}

\usepackage[misc,geometry]{ifsym}
\usepackage{listings}
\usepackage{graphicx}
\usepackage{xcolor}
\usepackage[colorlinks,
            linkcolor={black},
            citecolor={blue},
            urlcolor={blue},
            pdfauthor={No Author Given},
            bookmarks=false,
            pdfsubject={confidence-planner: Easy-to-Use Prediction
Confindence Interval Estimation},
            pdfkeywords={accuracy, confidence intervals, confidence bounds, cross-validation, bootstrapping, holdout}
            ]{hyperref}

\definecolor{darkgray}{rgb}{.4,.4,.4}

\begin{document}

\title{\texttt{confidence-planner}: Easy-to-Use Prediction Confidence Estimation and Sample Size Planning}

\author{Antoni Klorek\inst{1}\orcidID{0000-0001-5851-4045} \and Karol Roszak\inst{1}\orcidID{0000-0002-0786-5053} \and Izabela Szczech\inst{1}\orcidID{0000-0002-9655-4109} \and Dariusz Brzezinski\inst{1,2,3}\textsuperscript{(\Letter)}\orcidID{0000-0001-9723-525X}}
\authorrunning{A. Klorek, K. Roszak, I. Szczech, D. Brzezinski}
\titlerunning{\texttt{confidence-planner}: Easy-to-Use Prediction Confidence Estimation}

\institute{Institute of Computing Science, Poznan University of Technology, Poznan, Poland \and
Institute of Bioorganic Chemistry, Polish Academy of Sciences, Poznan, Poland  \and
MNM Bioscience, Cambridge, MA\\
\email{dariusz.brzezinski@cs.put.poznan.pl}}

\maketitle

\begin{abstract}
Machine learning applications, especially in the fields of me\-di\-cine and social sciences, are slowly being subjected to increasing scrutiny. Similarly to sample size planning performed in clinical and social studies, law-makers and funding agencies may expect statistical uncertainty estimations in machine learning applications that impact society. In this paper, we present an easy-to-use python package and web application for estimating prediction confidence intervals. The package offers eight different procedures to determine and justify the sample size and confidence of predictions from holdout, bootstrap, cross-validation, and progressive validation experiments. Since the package builds directly on established data analysis libraries it seamlessly integrates into preprocessing and exploratory data analysis steps. Code related to this paper is available at: \url{https://github.com/dabrze/confidence-planner}
\end{abstract}

\section{Introduction}
Medical, social, and behavioral sciences are known to be plagued by undersampling~\cite{SampleSizePlanner}. In the traditional statistical framework, even when the effect exists, undersampled studies yield either non-significant results or significant results because of overestimating the size of the effect. Similar problems can occur in machine learning studies on life-science data, where classification accuracy is often measured on relatively small samples without providing any uncertainty estimation. Importantly, statistical testing and uncertainty estimation of machine learning systems will become more and more common, as the global community realizes that AI systems need to be controlled to maintain fairness and comply with government regulations. Examples of such regulations have recently appeared in the proposed EU AI act~\cite{AI_act}, UNESCO recommendation on ethics in AI~\cite{unesco}, or the FDA's Software as Medical Device guidance~\cite{samd}.

To mitigate issues with undersampled studies, in social sciences authors are increasingly expected to plan and justify the sample size of their study. Such sample-size justification procedures can be found in tutorial papers and online tools~\cite{SampleSizePlanner}. Equivalent confidence interval and sample size estimation procedures for classification accuracy are scattered throughout scientific literature~\cite{DBLP:journals/jmlr/Langford05,DBLP:conf/colt/BlumKL99,Clopper-Pearson,bootstrap} and, to the best of our knowledge, are not available as a python package. The \texttt{confidence-planner} package presented herein aims to fill this gap.

\section{The \texttt{confidence-planner} Package}
The \texttt{confidence-planner} package provides implementations of estimation procedures for confidence intervals around classification accuracy. A \textit{confidence interval} (CI) is a range of estimates for an unknown parameter. In our case it is the range of values that we expect the accuracy $\mathit{acc}$ of our model to fall between if we re-run our experiment again. A confidence interval is computed for a test sample size $n$ and at a designated confidence level $\gamma$, which is the percentage of times one expects to reproduce an estimate between the upper and lower bounds of the confidence interval. A interval with a confidence level of 90\% is called a 90\% confidence interval (90\% CI).

For a given number of test samples $n$, test accuracy $\mathit{acc}$, and expected confidence level $\gamma$, \texttt{confidence-planner} offers a set of CI estimation procedures. The package currently features approximations for holdout (Langford~\cite{DBLP:journals/jmlr/Langford05}, Clopper-Pearson~\cite{Clopper-Pearson}, Wilson~\cite{Wilson}, Z-test, t-test), bootstrap~\cite{bootstrap}, cross-validation~\cite{DBLP:conf/colt/BlumKL99}, and progressive validation~\cite{DBLP:conf/colt/BlumKL99} experiments. Moreover, for selected methods, for a given confidence level $\gamma$, \texttt{confidence-planner} can help estimate the number of samples $n$ needed to obtain a CI of a user-specified radius. For example, using the Z-test approximation, \texttt{confidence-planner} can help estimate that in order to achieve a 90\% CI of $\pm 0.05$ one needs a holdout test of at least 271 examples.

The \texttt{confidence-planner} package is open source and available under a permissive MIT license. The source code, documentation and an introductory video are available at \url{https://github.com/dabrze/confidence-planner}. The
package can also be installed via PyPI using \texttt{pip install confidence-planner}. The package's basic functionality, along with guidance on selecting the appropriate estimation method, is also available the form of a confidence-planner web application that can be deployed using the code in the repository.

\section{Application Example}
Next, we present a basic application example featuring the well-known \textit{Breast Cancer Wisconsin} dataset to demonstrate how confidence interval estimation can be included in a data classification script. In this particular example, we will estimate the 90\% confidence interval (CI) of a classifier tested on a holdout test set using a Z-test approximation~\cite{DBLP:journals/jmlr/Langford05}. Then, we will estimate the holdout size that would be required to limit the 90\% CI to a radius 0.05 classification accuracy (estimated accuracy $\pm 0.05$). The complete code required to execute these tasks is the following:
\clearpage
\begin{lstlisting}[language=Python]
from sklearn import datasets, svm, metrics
from sklearn.model_selection import train_test_split
import confidence_planner as cp

# example dataset
X, y = datasets.load_breast_cancer(return_X_y=True)
X_train, X_test, y_train, y_test = train_test_split(
    X, y, test_size=0.3, stratify=y, random_state=23
)

# training the classifier and calculating accuracy
clf = svm.SVC(gamma=0.001)
clf.fit(X_train, y_train)
y_pred = clf.predict(X_test)
acc = metrics.accuracy_score(y_test, y_pred)

# confidence interval and sample size estimation
ci = cp.estimate_confidence_interval(
    len(y_test), acc, 0.90, method="holdout_z_test"
)
sample = cp.estimate_sample_size(0.05, 0.90, method="holdout_z_test")
print(f"Holdout accuracy: {acc}")
print(f"90% CI: {ci}")
print(f"Test samples needed for a 0.05 radius 90% CI: {sample}")
\end{lstlisting}

The first three lines import the sklearn library used for classification and the
\texttt{confidence-planner} package. The following fragments of code load the data into a standard pandas DataFrame object, split the data into a training and a holdout test set, train an SVM classifier and record its classification accuracy on the holdout. The final fragment of code performs the 90\% CI estimation and calculates the number of samples that would be needed to make CI radius equal 0.05. Similar estimations can be performed for cross-validation, bootstrapping, and progressive validation, by specifying a different estimation function.

\begin{figure}[!htb]
	\centering
    \includegraphics[width=0.82\textwidth]{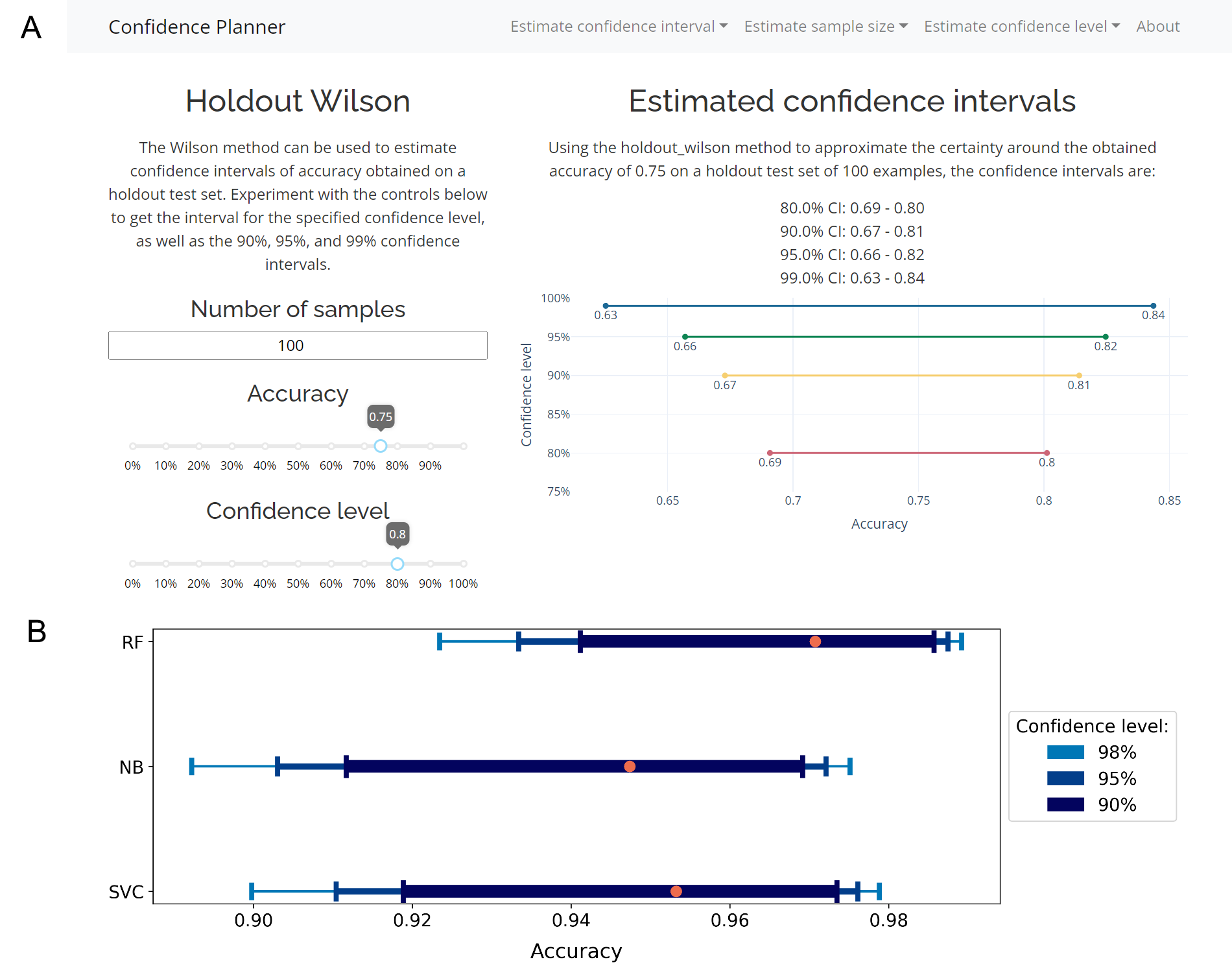}
    \caption{Confidence-planner screenshots. \textbf{A}) Screenshot of the confidence planner web application. \textbf{B}) Graded error bars produced by the confidence-planner package.}
    \label{fig:planner_app}
\end{figure}

The same analyses can be performed online, without coding, by using the confidence-planner web app. Figure~\ref{fig:planner_app} shows the CI estimation page for the Wilson method for holdout data and graded error bars that can be created using the python package.

\section{Conclusions}
This demo paper introduced the \texttt{confidence-planner} package that enables calculating confidence bounds around classification accuracy. It provides an easy-to-use, extensible, and freely available implementation of estimation procedures for holdout, bootstrap, cross-validation and progressive validation schemes. In the future we plan to extend the list of estimation methods and provide more visualizations of uncertainty, for example for validation and learning curves.

\bibliographystyle{splncs04}
\bibliography{ConfidencePlanner}

\end{document}